\documentclass[prd,preprint,tightenlines,floatfix,showpacs,
preprintnumbers,nofootinbib,eqsecnum]{revtex4-1}

\usepackage[T1]{fontenc}		

\usepackage{amsmath,amsfonts,amssymb,amstext,mathrsfs}
\usepackage{mathpazo}

\usepackage[dvips]{graphicx}
\usepackage{epsf,float}
\usepackage{revsymb}

\usepackage{dcolumn}
\usepackage{braket}
\usepackage{color,xcolor}
\usepackage{graphicx}
\usepackage{subfigure}
\usepackage{multirow}
\usepackage{tabularx}
\usepackage{pstricks}
\usepackage[section]{placeins}
\usepackage{booktabs}
\usepackage{array}

\usepackage{hyperref}

\newcommand{\KaTie}{{\sc Ka\nolinebreak\hspace{-0.3ex}Tie}}

\begin{document}

\title{Single- and double-scattering production of four muons\\ in ultraperipheral $PbPb$ collisions 
at the Large Hadron Collider}

\author{Andreas van Hameren}
\email{hameren@ifj.edu.pl}
\author{Mariola K{\l}usek-Gawenda}
\email{Mariola.Klusek@ifj.edu.pl}
\author{Antoni Szczurek\footnote{Also at Faculty of Mathematics and Natural Sciences, University of Rzeszow, ul. Pigonia 1, 35-310 Rzesz\'ow, Poland.}}
\email{Antoni.Szczurek@ifj.edu.pl}
\affiliation{Institute of Nuclear Physics, Polish Academy of Sciences, Radzikowskiego 152,
PL-31-342 Krak\'ow, Poland}

\date{\today}

\vspace{2cm}

\begin{abstract}
We discuss production of two
$\mu^+\mu^-$ pairs in ultraperipheral ultrarelativistic heavy 
ion collisions at the LHC.
We take into account electromagnetic (two-photon) double-scattering
production and for a first time direct $\gamma\gamma$ production of 
four muons in one scattering.
We study the unexplored process
$\gamma \gamma \to \mu^+\mu^-\mu^+\mu^-$.
We present predictions for total and differential cross sections.
Measurable nuclear cross sections are obtained and corresponding differential distributions and
counting rates are presented.
\end{abstract}



\maketitle

\section{Introduction}

Ultraperipheral heavy ion scattering is a special category 
of nuclear collisions \cite{Budnev:1974de,Bertulani:1987tz,Baur:2001jj,Bertulani:2005ru,Baltz:2007kq}.
This field got a new impulse with the start of the LHC.
Several final states are possible in ultraperipheral heavy ion
collisions (UPC). Charged particles are final states that 
are relatively easy to measure.
The present experiments on UPC concentrated rather on two-body final states.
Typical examples are $e^+ e^-$ \cite{Klusek-Gawenda:2016suk}, $\mu^+ \mu^-$ \cite{KGSz:muons}, 
$\pi^+ \pi^-$ \cite{Klusek-Gawenda:2013rtu}. $p \bar p$ scattering is another interesting
possibility \cite{KLNS2017}. Recently the ATLAS collaboration has 
demonstrated that also $\gamma \gamma$ final states are available
\cite{Aaboud:2017bwk} leading to a first experimental verification
of elastic $\gamma \gamma \to \gamma \gamma$ scattering (for a theoretical
work see e.g. \cite{Klusek-Gawenda:2016euz}).

In principle, also four-body final states can be measured.
For example the STAR collaboration obtained some results for
$\pi^+ \pi^- \pi^+ \pi^-$ \cite{Abelev:2009aa}. Here both resonant and nonresonant
contributions can be present \cite{Klusek-Gawenda:2013dka}. Here the double-scattering
(two subsequent $\gamma \gamma$ interactions) in UPC
may play an important role.

Recently we made a first estimation of the production of
four electrons\footnote{Charge conservation is understood. We will regularly refer to ``electrons'' and ``muons'' where we mean neutral combinations of electron-positron pairs and muon-antimuon pairs.} in UPC of two lead ions \cite{Klusek-Gawenda:2016suk}.
Only the double-scattering contribution was considered
there. Here we do similar calculation but for four muon production.
However, in addition to the double-scattering mechanism we include here for a first time
also the contribution of the single-scattering process with underlying
$\gamma \gamma \to \mu^+ \mu^- \mu^+ \mu^-$ elementary process.
We wish to explore the competition of the two mechanisms.
The single-scattering mechanism in nuclear collisions was not discussed so far in the literature.
Therefore we start from considering the elementary
$\gamma \gamma \to \mu^+ \mu^- \mu^+ \mu^-$ subprocess, also not
discussed in the literature.

\section{Some theoretical aspects of muon production 
in heavy ion UPC }

\begin{figure}[!h]
(a) \includegraphics[scale=0.35]{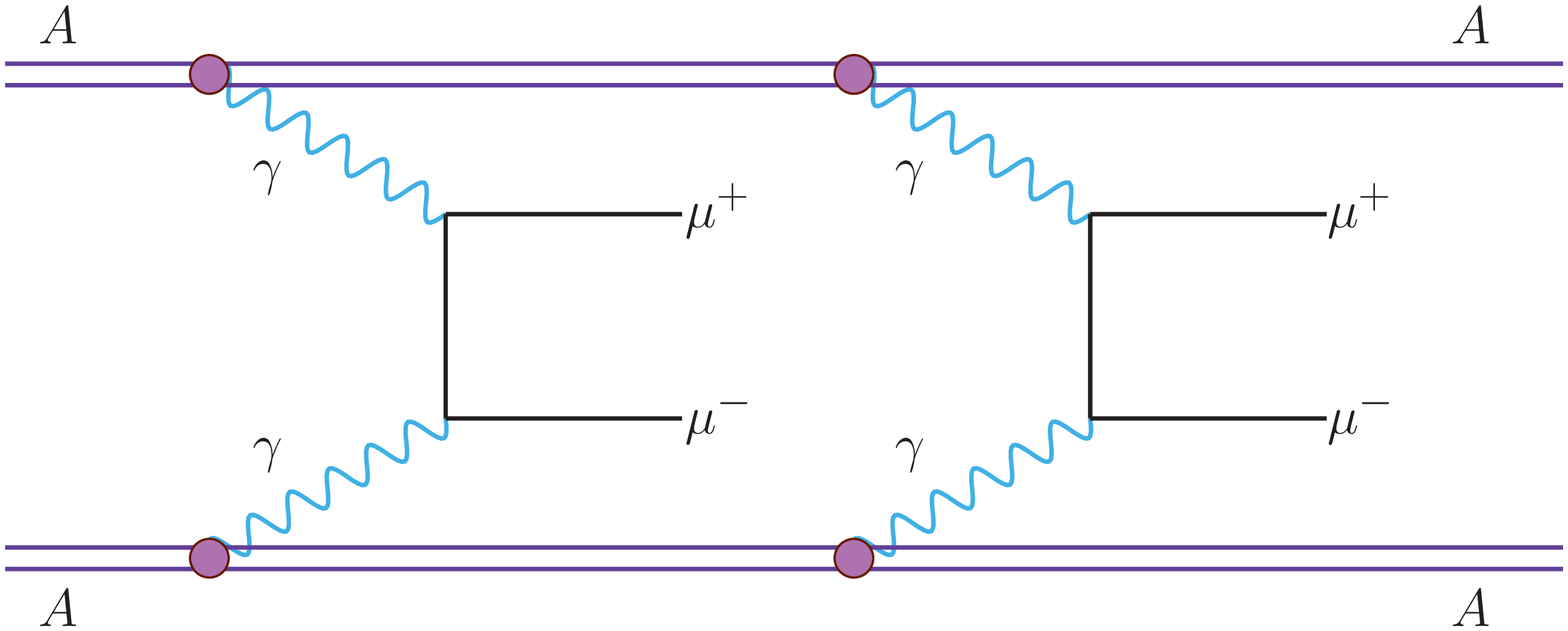}
(b) \includegraphics[scale=0.35]{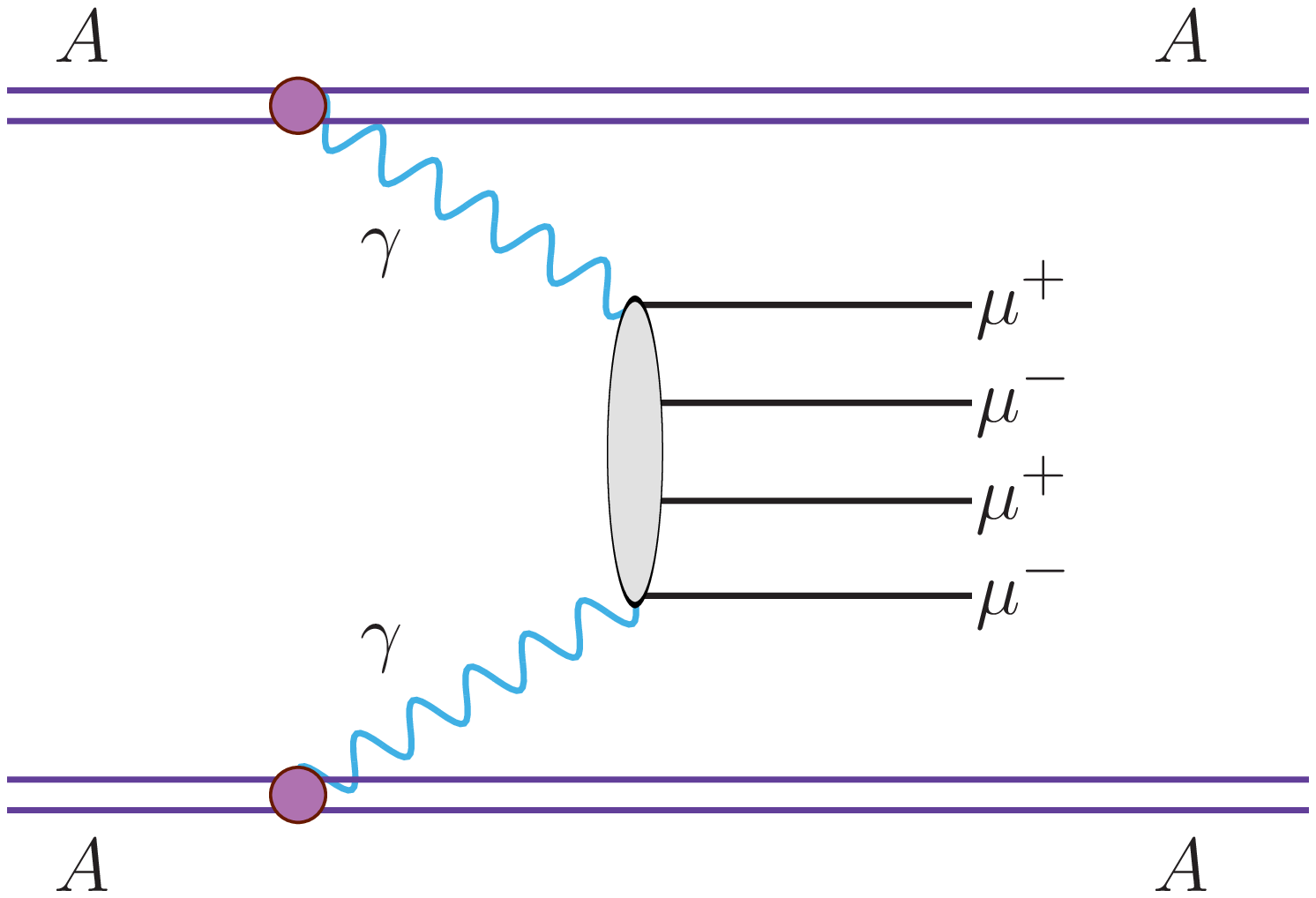}
\caption{Diagrams for 
(a) double-scattering and 
(b) direct (single-scattering) 
production of two $\mu^+\mu^-$ pairs in ultrarelativistic UPC of 
heavy ions.}
\label{fig:diagrams}
\end{figure}

The differential cross section for double-scattering production of four muons 
can be expressed through a probability density to produce a first (superscript $I$) 
and second (superscript $II$) $\mu^+\mu^-$ pair 
\begin{eqnarray}
\frac{{\rm d} \sigma_{AA \to AA \mu^+\mu^-\mu^+\mu^-}}{	{\rm d} y_{\mu^+} {\rm d} y_{\mu^-} {\rm d} p_{t,\mu}
						{\rm d} y_{\mu^+} {\rm d} y_{\mu^-} {\rm d} p_{t,\mu}} =  
						\frac{1}{2} \int {\rm d}^2 b 
						& \times &
\frac{{\rm d} P^I_{AA ^{\underrightarrow{\gamma\gamma}} AA \mu^+\mu^-} \left(b; y_{\mu^+}, y_{\mu^-}, p_{t,\mu} \right)}{{\rm d} y_{\mu^+} {\rm d} y_{\mu^-} {\rm d} p_{t,\mu}}  \nonumber \\  
&\times&
\frac{{\rm d} P^{II}_{AA ^{\underrightarrow{\gamma\gamma}} AA \mu^+\mu^-} \left(b; y_{\mu^+}, y_{\mu^-}, p_{t,\mu} \right)}{{\rm d} y_{\mu^+} {\rm d} y_{\mu^-} {\rm d} p_{t,\mu}} \;.
\label{eq:diff_sigma_4mu}
\end{eqnarray}
The additional factor $1/2$ comes from identity of the two pairs.
The probability for each of pair is the same, but the correlations between muons
of the same sign are different than correlations for an opposite-sign pair. 
Above $b$ is an impact parameter i.e.\ the distance between colliding nuclei
in the transverse direction,
$y_\mu$ is the rapidity of a muon and $p_{t,\mu}$ is transverse momentum
of a muon. Here we use the fact that both muons in a given scattering 
(both first and second scattering) have the same transverse momenta.  
The formula for the differential cross section (Eq.(\ref{eq:diff_sigma_4mu})) 
allows to control (calculate) the whole kinematics of the outgoing particles 
(e.g.\ scattering angle of each of muons and invariant mass 
of two/four muons).
The probability for the production of the two muon pairs reads:
\begin{eqnarray}
P_{AA ^{\underrightarrow{\gamma\gamma}} AA \mu^+\mu^-} \left(b; y_{\mu^+}, y_{\mu^-}, p_{t,\mu} \right) &=& 
\int \frac{{\rm d} \sigma_{\gamma\gamma \to \mu^+\mu^-} \left(W_{\gamma\gamma}\right)}{{\rm d}z}  
N\left( \omega_1, \textbf{b}_1 \right) N\left( \omega_2, \textbf{b}_2 \right) 
S^2_{abs}\left( \textbf{b} \right) \nonumber \\
& \times & \frac{W_{\gamma\gamma}}{2}
{\rm d} W_{\gamma \gamma} \, {\rm d} Y_{\mu^+\mu^-} \, {\rm d} z \, {\rm
  d} \overline{b}_x \, {\rm d} \overline{b}_y  \; .
\label{eq:prob_byypt}
\end{eqnarray}
The cross section for the production of a single muon pair in heavy-ion collisions
\begin{eqnarray}
\sigma_{AA \to AA \mu^+\mu^-} = \int {\rm d}^2 b \times \frac{{\rm d} P_{AA ^{\underrightarrow{\gamma\gamma}} AA \mu^+\mu^-} \left(b; y_{\mu^+}, y_{\mu^-}, p_{t,\mu} \right)}{{\rm d} y_{\mu^+} {\rm d} y_{\mu^-} {\rm d} p_{t,\mu}}
\times {\rm d} y_{\mu^+} {\rm d} y_{\mu^-} {\rm d} p_{t,\mu} \;.
\end{eqnarray}  
In Eq.~(\ref{eq:prob_byypt}) $W_{\gamma\gamma}=\sqrt{4\omega_1\omega_2}$ 
is the energy available in the $\gamma\gamma$ system and 
$\omega_i$ is energy of the photon which is emitted
from the first or second nucleus. 
$Y_{\mu^+\mu^-}=\frac{1}{2}(y_{\mu^+} + y_{\mu^-})$
is the  rapidity of the outgoing muon pair. 
The quantities $\overline{b}_x$, $\overline{b}_y$
are the components of the $\overline{{\bf b}} = (\textbf{b}_1 + \textbf{b}_2)/2$ vector
where $\textbf{b}_1$ and $\textbf{b}_2$ indicate the point (transverse distance from the first and second nucleus) 
in which the photons collide with each other and particles (muons) are produced. 
A diagram illustrating the quantities in the impact parameter
space can be found for example in Ref.~\cite{KGSz:muons}. 
In our calculations we use the so-called realistic form factor 
which is a Fourier transform of the charge distribution in the nucleus. 
A detailed study of this form factor and its role was presented e.g.\ in Ref.~\cite{KGSz:muons}.
In the same paper an expression for the photon flux    
$N\left( \omega_i, \textbf{b}_i \right)$ and its
dependence on the charge form factor of the ``emitting'' nucleus was given explicitly.

In the case of direct four-muon production we use the following formula:
\begin{equation}
\sigma_{4 \mu} = \int \sigma_{\gamma \gamma \to 4 \mu}(W_{\gamma
  \gamma})
N(\omega_1,b_1) N(\omega_2,b_2) S_{abs}^2(b) d^2b d \omega_1 d \omega_2
d \overline{b}_x d \overline{b}_y   \;  .
\label{nuclear_4mu}
\end{equation}
By appropriate binning one can obtain the distribution in 
$M_{4 \mu} = W_{\gamma \gamma}$.

\subsection{Dimuon production in Pb+Pb UPC}

\begin{figure}
	\includegraphics[scale=0.4]{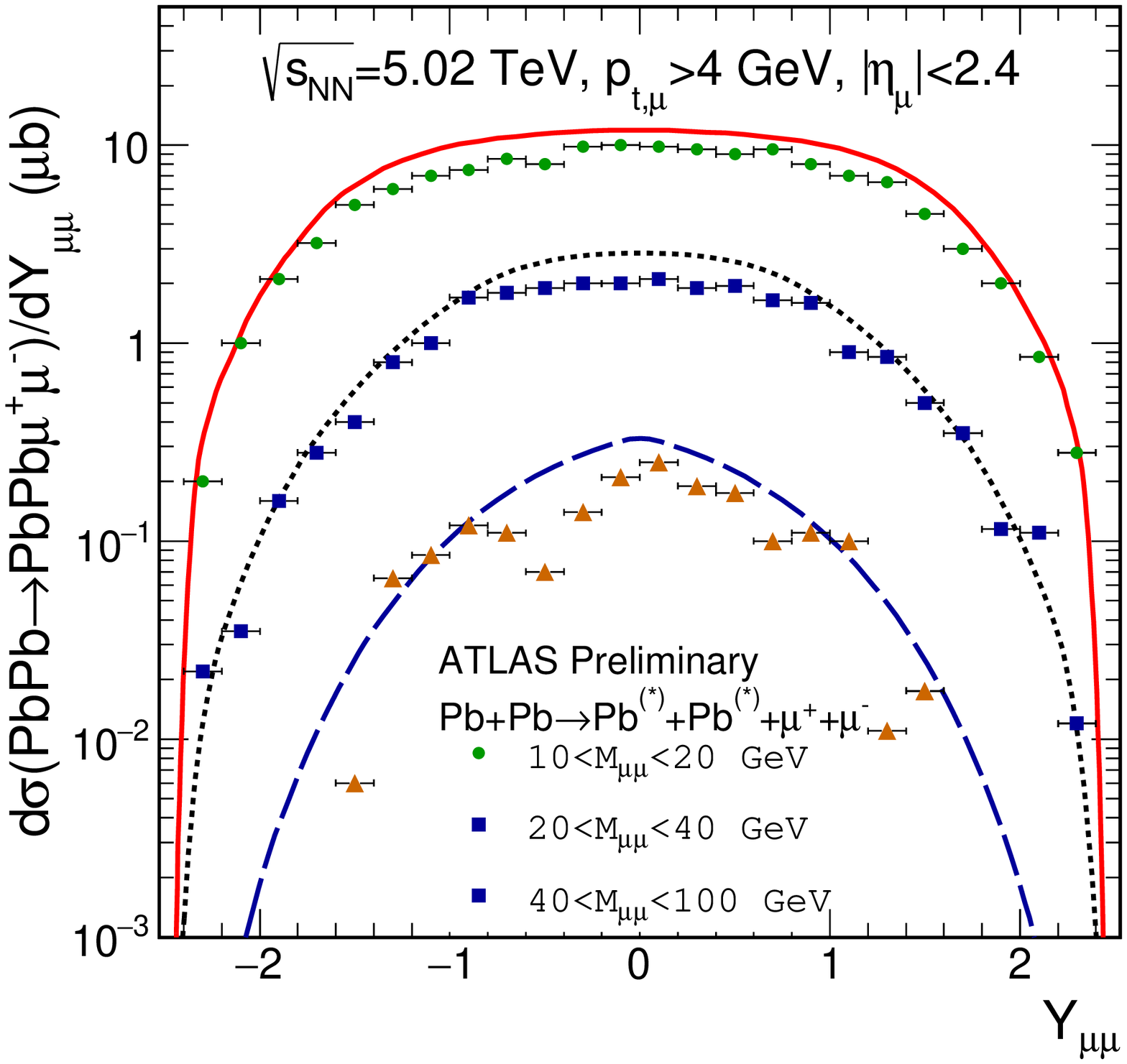}
	\includegraphics[scale=0.4]{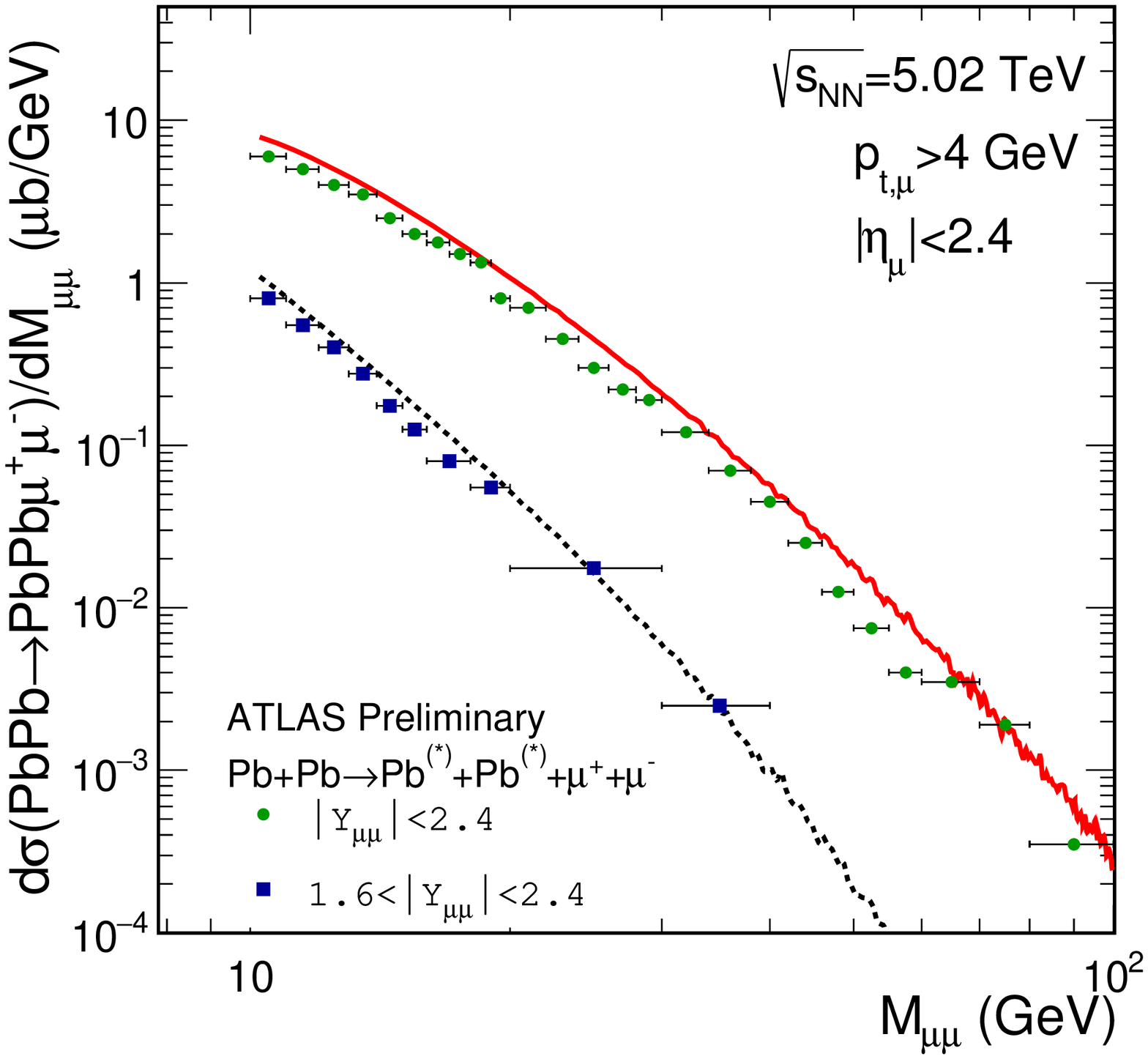}
	\caption{Nuclear cross section as a function of dimuon rapidity (left panel) and dimuon invariant mass (right panel). Theoretical predictions are compared with preliminary
	ATLAS data scanned by us from figures in \cite{ATLAS:2016vdy}.}
	\label{fig:ATLAS_UPC}
\end{figure}

Before going to the production of two (neutral) dimuon pairs we wish to check
for the first time our predictions for the production of one muon pair with
the preliminary data of the ATLAS collaboration~\cite{ATLAS:2016vdy}.
In our calculations we imposed cuts on muon rapidities
-2.4 $< y_i <$ 2.4 and on muon transverse momenta $p_{t,i} >$ 4 GeV.
In Fig.~\ref{fig:ATLAS_UPC} we show distributions in the rapidity of the pair
$Y_{\mu \mu}$ (left panel) for different windows of dimuon invariant
mass specified in the figure and in dimuon invariant mass (right panel) 
for two windows of $Y_{\mu \mu}$ specified in the figure.
We get slightly larger cross sections than observed experimentally.
There can be a few reasons. Perhaps not all experimental
conditions were included by us. In our present simplified treatment we
assume $S_{abs}^2 = \theta \left( b - |R_1 + R_2| \right)$ which is
a rather crude approximation for collisions when peripheries of nuclei
interact with each other. We think this is, however, a sufficiently precise
approximation for our first estimation of four-muon production in UPC.
A slightly better agreement was obtained in~\cite{Klusek-Gawenda:2016suk} for the reaction PbPb$\to$PbPb$e^+e^-$
measured by the ALICE collaboration \cite{Abbas:2013oua}.

\section{$\gamma \gamma \to \mu^+\mu^-\mu^+\mu^-$ elementary 
cross section}

In our approach, the cross section for the elementary $\gamma\gamma \to
\mu^+\mu^-$ process is one of the basic ingredients for the nuclear 
double-scattering mechanism.
The elementary process with on-shell photons could not yet be studied
experimentally. It was demonstrated recently that in nuclear collisions the subprocess
$\gamma\gamma \to \mu^+\mu^-$ can be observed~\cite{ATLAS:2016vdy}. 
The total cross section for $\gamma \gamma \to \mu^+ \mu^-$ is 
well known \cite{Budnev:1974de} and can be obtained
based on standard Feynman diagram technique.
In our calculations of one pair production
as well as the double parton scattering 
we use rather differential cross section 
as a function of $z=\cos \theta$, where $\theta$ is muon scattering angle.
The corresponding formula is also well known:
\begin{eqnarray}
\frac{{\rm d} \sigma}{{\rm d} z} = \frac{\left(4 \pi \alpha_{em} \right)^2 v}{8 \pi W^2}
\frac{1+2 v^2 (1-v^2)(1-z^2)-v^4 z^4}{(1-v^2 z^2)^2} \; ,
\end{eqnarray}      
where $v = \sqrt{1- \frac{4m_\mu^2}{W^2}}$.

\begin{figure}[!h]
\includegraphics[scale=0.4]{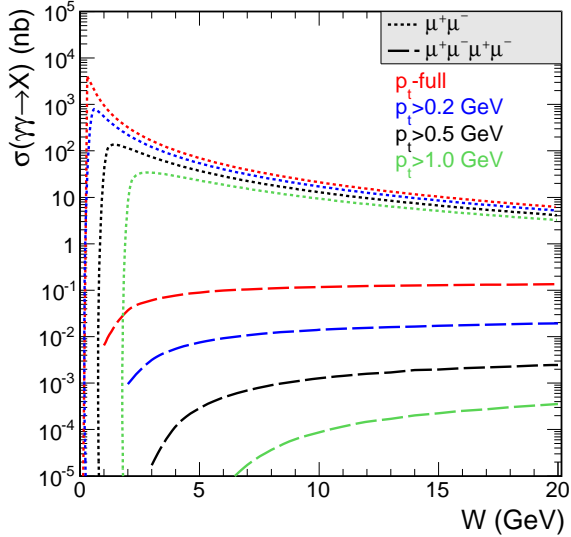}
\caption{Total cross section for $\gamma \gamma \to \mu^+ \mu^-$ (dotted
  lines) and $\gamma \gamma \to \mu^+ \mu^- \mu^+ \mu^-$ (dashed lines) 
as a function of $\gamma\gamma$ collision energy 
for different cuts on transverse momenta of all muons
in the final state.}
\label{fig:sigma_W}
\end{figure}

The calculations for $\gamma\gamma \to \mu^+\mu^-\mu^+\mu^-$ were performed with the help of
\KaTie~\cite{vanHameren:2016kkz}. It is an event generator that is
specially designed to deal with initial states that have an explicit
transverse momentum dependence, but can also deal with on-shell initial
states, like the ones we consider here. Furthermore, \KaTie\ is a parton-level
generator for hadron scattering, but requires only a few adjustments to
deal with photon scattering. All necessary tools to sample the phase
space and to calculate the scattering amplitude are available in the
library {\sc avhlib}~\cite{Bury:2015dla}, which \KaTie\ employs.
Amplitudes for processes with several final-state particles are
calculated numerically via recursive methods keeping the computational
complexity under control. Large algebraic expressions are avoided, and
the numerical approach suffices for the generation of event files, which
can then be used to obtain the distributions of the desired variables.

Fig.~\ref{fig:sigma_W} presents the elementary cross section 
for the production of $\mu^+\mu^-$ (dotted lines) and $\mu^+\mu^-\mu^+\mu^-$
(dashed lines). Here we consider the single-scattering mechanism only.
This means that muons are produced directly from $\gamma\gamma$ fusion.
We show the total cross section as a function of energy for four different ranges
of transverse momentum ($p_{t,\mu}$ in full range, 
$p_{t,\mu}$ larger than $0.2$, $0.5$ and $1$ GeV). 
We see that the value of $p_{t,\mu}^{min}$ has an influence mainly on small 
$W_{\gamma\gamma}$ for $\gamma\gamma \to \mu^+\mu^-$.
For four-muon production, the larger value of $p_{t,\mu}^{min}$ the smaller
the cross section over the whole range of energy.   
Simultaneously, it is important to note that the cross section for single muon
pair production is at least one order of magnitude larger at $W=20$ GeV, and that it is about four orders of magnitude larger than the cross section for four-muon production for smaller values of $W$ over the whole range of $p_{t,\mu}$.
In contrast to two-muon production, the cross section for four-muon pair 
production increases with larger $W$.

\begin{figure}[!h]
\includegraphics[scale=0.4]{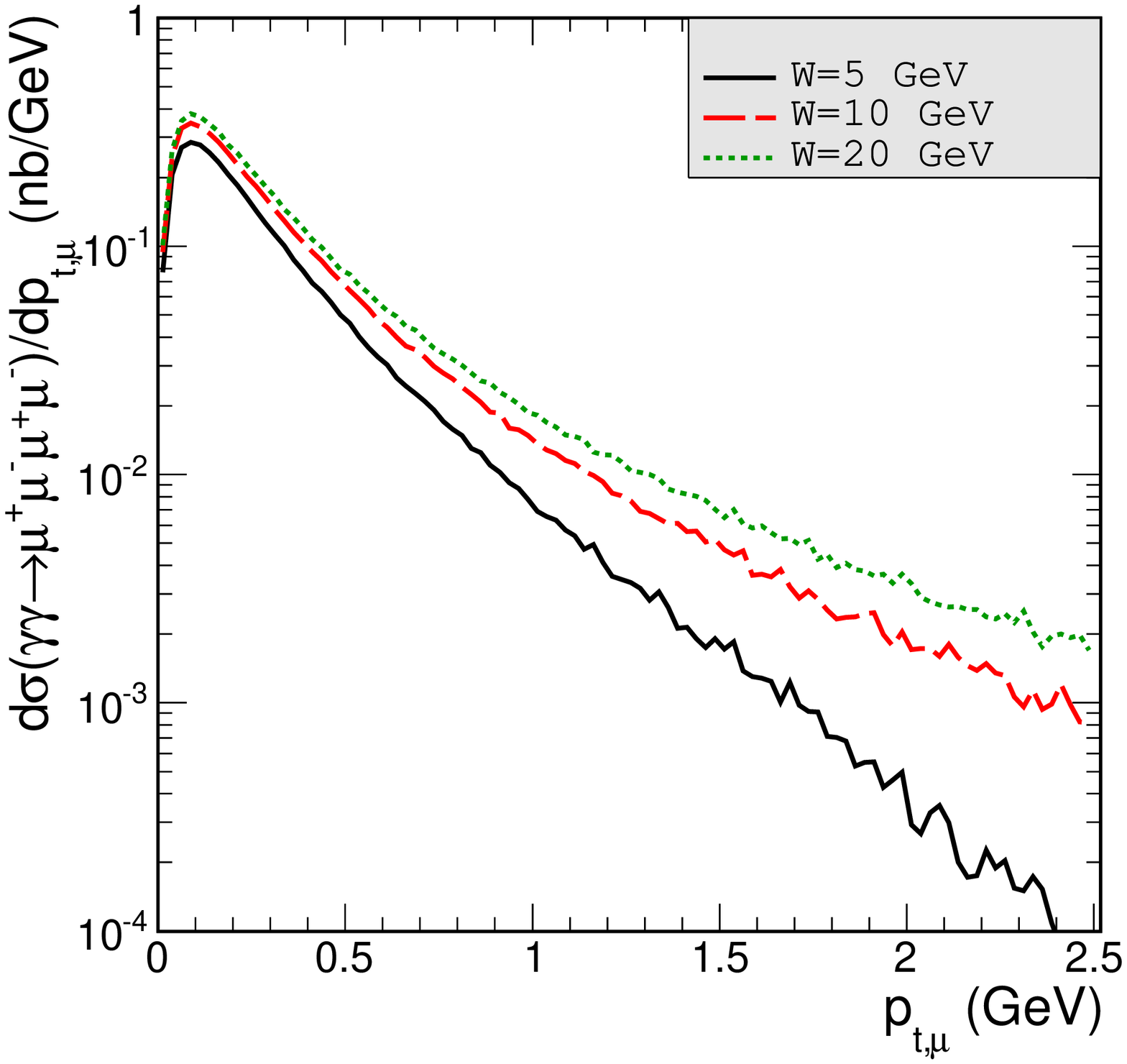}
\includegraphics[scale=0.4]{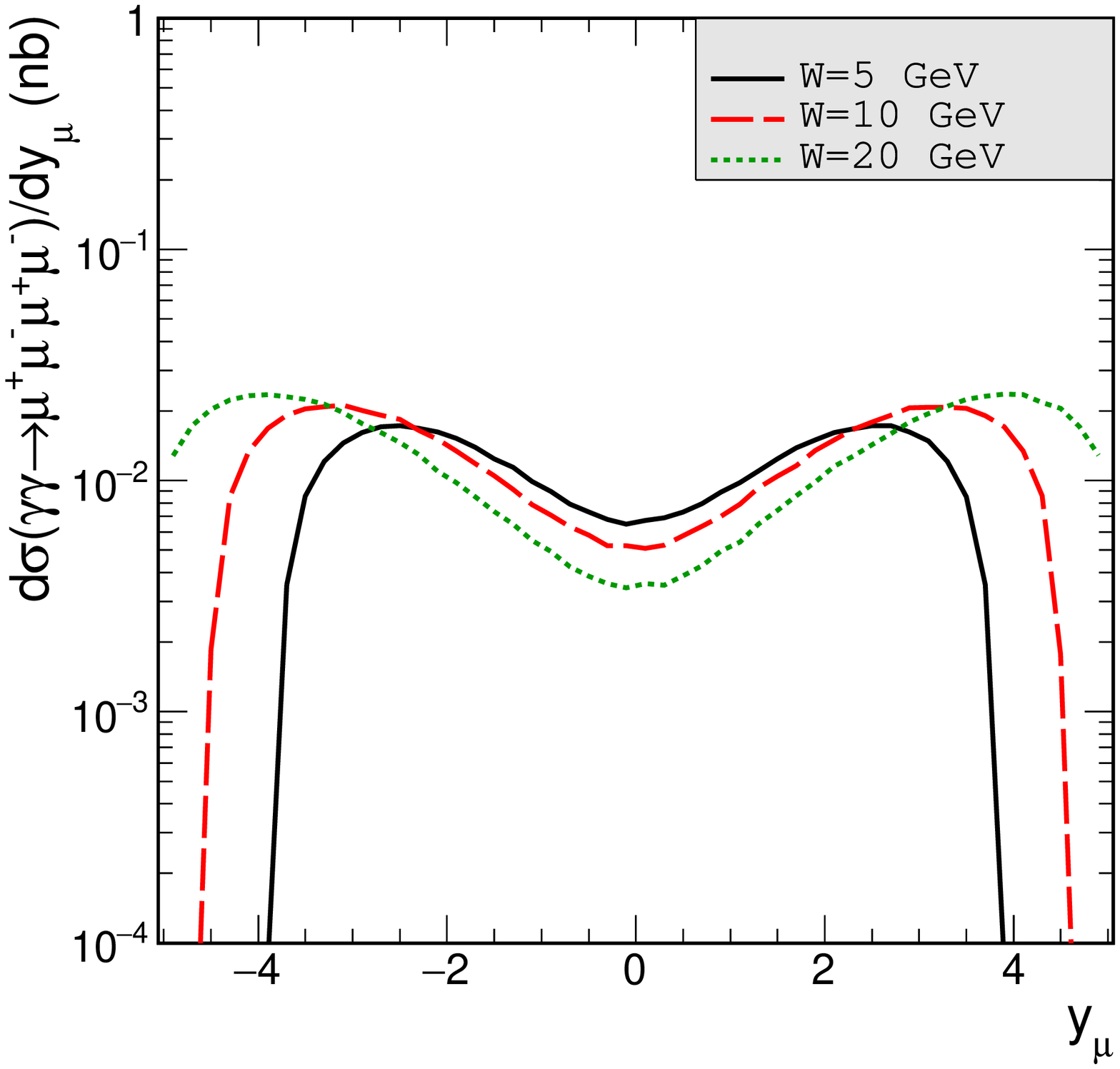}
\caption{Differential cross section as a function of transverse momentum
  of muon (left panel) and muon rapidity (right panel) for 
different values of the $\gamma\gamma$ collision energy specified in the figure.}
\label{fig:dsig_SS}
\end{figure}

Direct production of two $\mu^+\mu^-$ pairs is a new mechanism which
was never considered before. It seems interesting to examine this 
reaction more closely. 
We would like to study kinematic characteristics of one muon or one pair
of muons out of four muons in the final state.
The left panel of Fig.~\ref{fig:dsig_SS} shows the differential cross section 
as function of the transverse momentum of one randomly selected muon. The three lines present 
results for three different values of the $\gamma\gamma$ collision energy 
($W=5$, $10$ and $20$ GeV).
Obviously, the larger energy the larger the total cross section.
The right panel of the same figure presents the differential cross section
as a function of the muon rapidity. In general, muons 
in the four-body final state are produced 
in the forward/backward directions rather than in the mid-rapidity range.

\begin{figure}[!h]
	\includegraphics[scale=0.4]{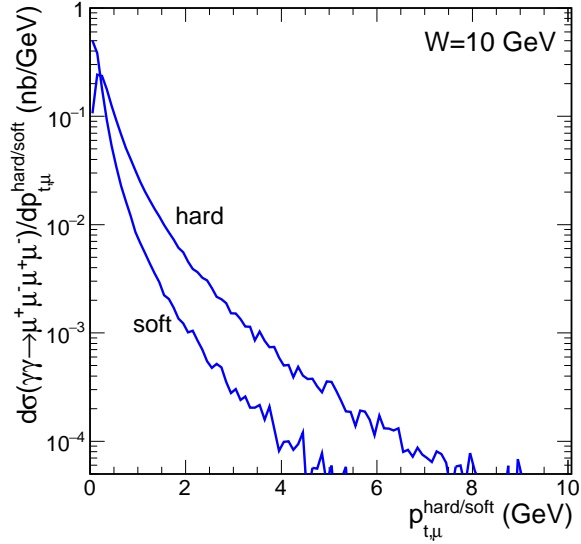}
	\caption{Differential cross section as a function of
	hard and soft muon transverse momentum.}
\label{fig:dsig_dpt_hard_soft}
\end{figure}

Experimentally, some times, it is more useful to make distributions of
harder and softer muons separately for positive and negative ones separately. 
By harder we mean the muon with $p_t = \max\{p_{1t},p_{3t}\}$
	(or $p_t = \max\{p_{2t},p_{4t}\}$) and by softer we mean the muon with $p_t = \min\{p_{1t},p_{3t}\}$ (or $p_t = \min\{p_{2t},p_{4t}\}$).
In Fig.~\ref{fig:dsig_dpt_hard_soft} we show an example for the $\gamma \gamma$ collision energy W = 10 GeV.

\begin{figure}[!h]
\includegraphics[scale=0.4]{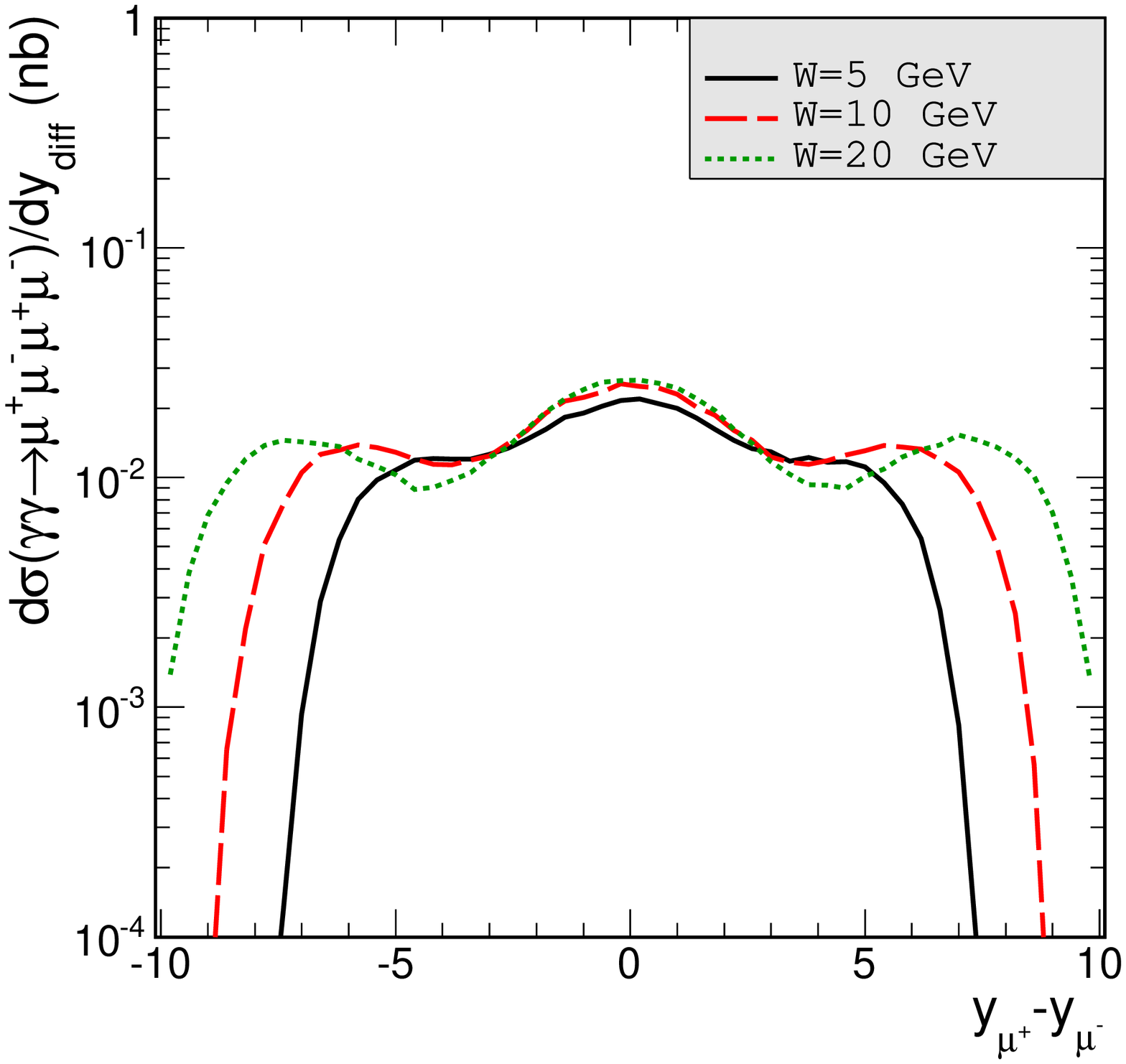}
\includegraphics[scale=0.4]{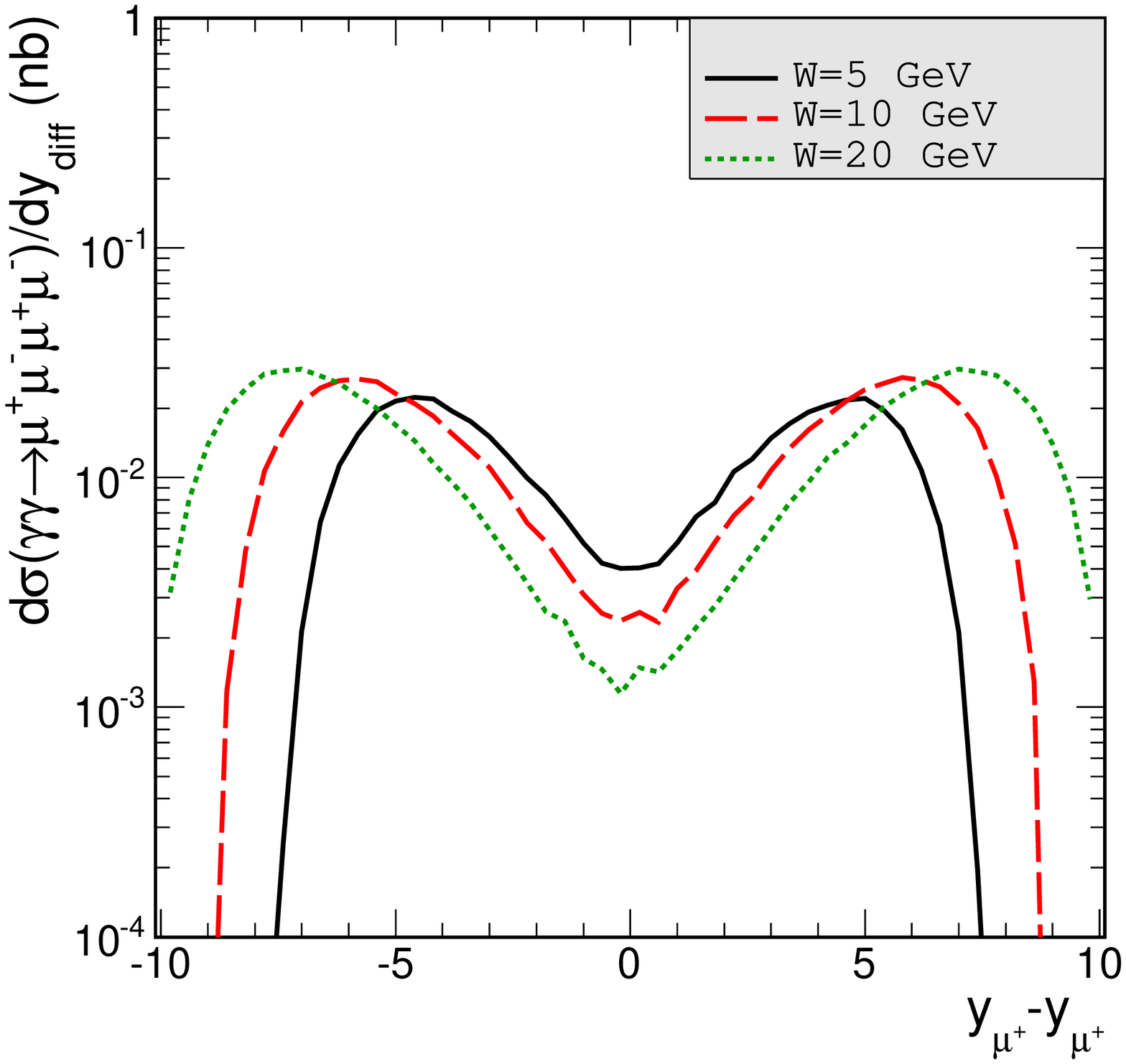}
\caption{Differential cross section as a function of rapidity difference
between the opposite-sign muons (left panel) and
between the same-sign muons (right panel) 
for the $\gamma \gamma \to 4 \mu$ reaction. 
The three lines in the figure relate to different values of 
the $\gamma\gamma$ collision energy given explicitly in the figure.}
\label{fig:dsig_dydiff_SS}
\end{figure}

Fig.~\ref{fig:dsig_dydiff_SS} shows the differential cross section
as function of the rapidity difference for different combinations of muons. The left panel depicts the distribution in 
the difference of the rapidities of opposite-sign muons 
and the right panel is for the same-sign muons.
We see the largest probability for the case when muons of the opposite-sign 
fly both at small and large rapidity distances
while muons of the same-sign fly preferably at large rapidity distances. 

\section{Predictions for four muons production 
in UPC of heavy ions}

At present the process 
$\gamma \gamma \to \mu^+ \mu^- \mu^+ \mu^-$  cannot be studied experimentally.
This process can be, however, studied in collisions of heavy ions.
In the present calculation we assume $\sqrt{s_{NN}}$ = 5.02 TeV.

\begin{figure}[!h]
\includegraphics[scale=0.4]{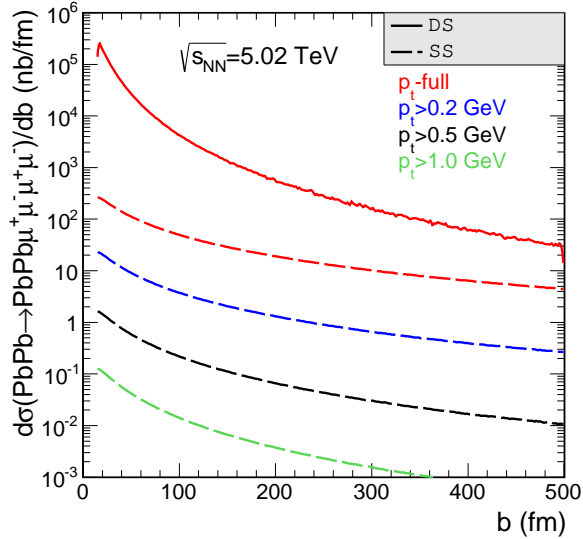}
\caption{The differential cross section for 
PbPb$\to$PbPb$\mu^+\mu^-\mu^+\mu^-$ 
as function of the impact parameter.
The top solid line  corresponds to double-scattering contributions
while the dashed lines below are for single scattering.
Each of the dashed lines corresponds to a distinct cut on transverse momenta of each muon.
}
\label{fig:dsig_dbm}
\end{figure}

In Fig.~\ref{fig:dsig_dbm} we show the (rather academic) distribution of the impact
parameter for lead-lead collisions.
Of course such a distribution cannot be measured experimentally. 
The solid line on top corresponds to double-scattering mechanism. 
No cuts on rapidities nor on muon transverse momenta are included here.
The contribution of single-scattering is shown by the dashed lines
below, for different cuts on the muon transverse momenta
specified in the figure.
The single-scattering cross sections are smaller than the double-scattering 
contribution. This confirms our naive expectations in \cite{Klusek-Gawenda:2016suk}.

\begin{figure}
\includegraphics[scale=0.4]{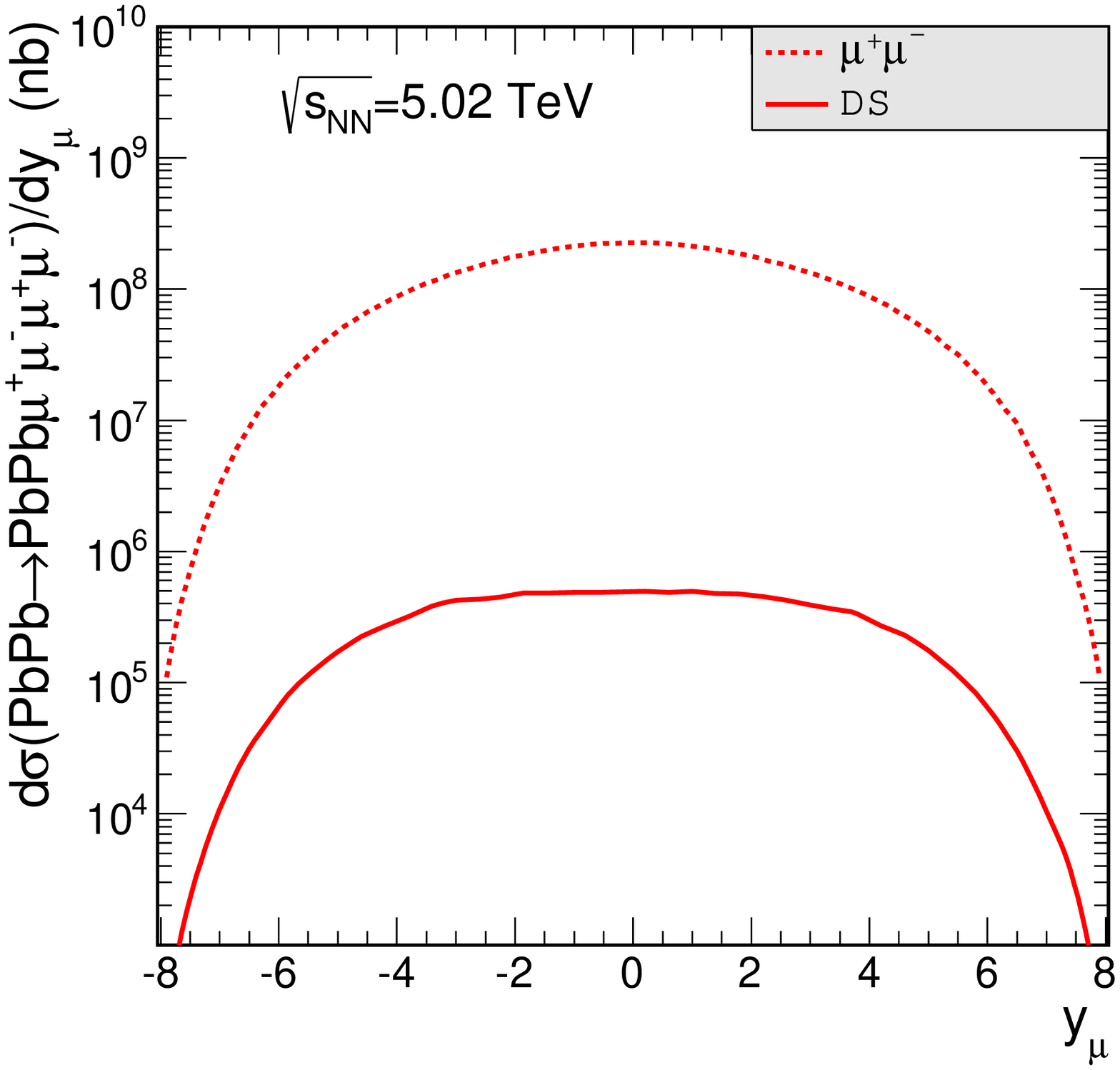}
\includegraphics[scale=0.4]{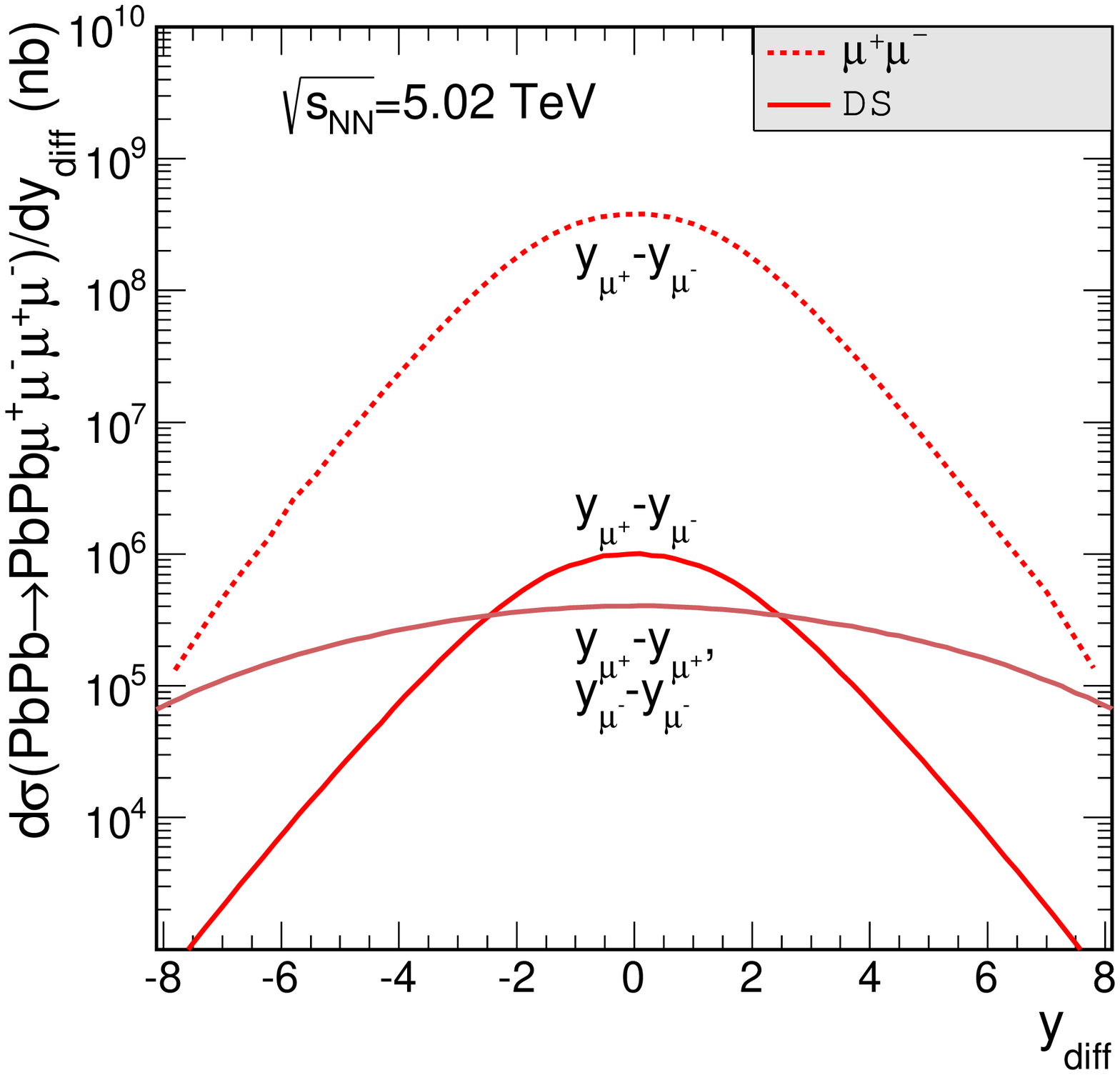}
\caption{Differential cross section for the PbPb$\to$PbPb$\mu^+\mu^-\mu^+\mu^-$
reaction as function of the
rapidity of a single muon $d\sigma/dy_{\mu^+_1} = d\sigma/dy_{\mu^+_2} =
d\sigma/dy_{\mu^-_1} = d\sigma/dy_{\mu^-_2}$ (left panel)
and the rapidity difference of selected types of muons (right panel)
for the double-scattering mechanism.}
\label{fig:dsig_dy}
\end{figure}

In the left panel of Fig.~\ref{fig:dsig_dy} we show rapidity distributions of muons from
the double-scattering mechanism 
(dashed line at the bottom) as well as for 
the single $\mu^+ \mu^-$ pair production (top dashed line).
Clearly the cross section for the two-pair production is more than three
orders of magnitude smaller than that for the single-pair production.
In the right panel we show corresponding differential distributions in several
differences of rapidities of the produced muons. 
The muons from the same scattering are strongly correlated.
The corresponding distribution peaks at $y_{diff} =$ 0.
There is no similar correlation for muons produced in different
scattering (the same-sign muons).
It is not easy to show similar distributions for single-scattering
with the four-body final state. This can be understood by inspecting
the organization of our nuclear calculations, where first a simple
energy dependent grid of the cross section for $\gamma \gamma \to 4 \mu$ is prepared
with the help of the Monte Carlo code \KaTie\ which is then used for calculating the 
nuclear cross section (see Eq.~\ref{nuclear_4mu}). Preparing more detailed multi-dimensional 
distributions that could be used in nuclear calculations clearly goes beyond
the scope of the present paper.

\begin{figure}
\includegraphics[scale=0.4]{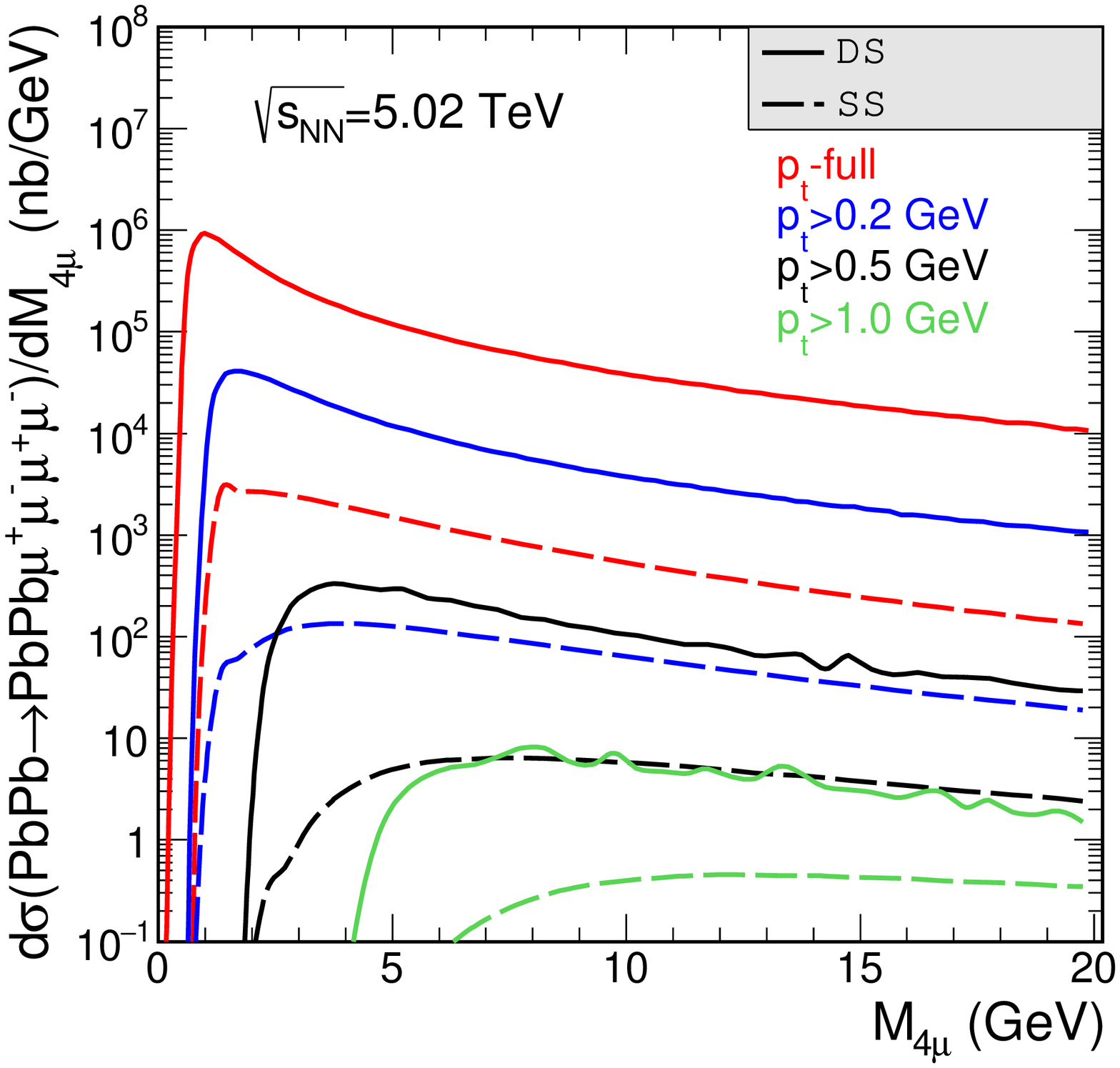}
\includegraphics[scale=0.4]{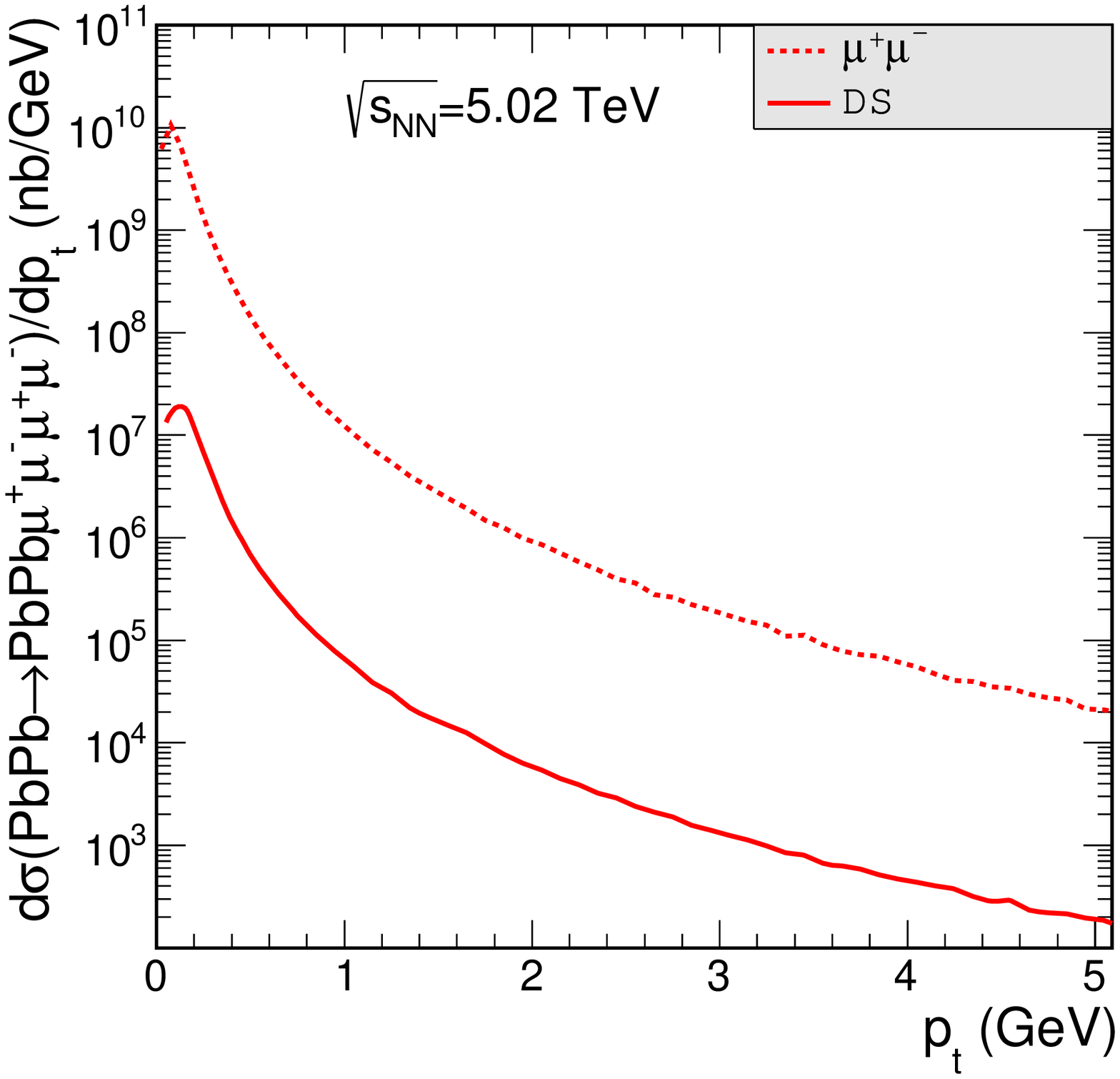}
\caption{Differential cross section for PbPb$\to$PbPb$\mu^+\mu^-\mu^+\mu^-$ 
as function of the invariant mass of four muons (left panel) and
transverse momentum of each of the muons (right panel).}
\label{fig:dsig_dW}
\end{figure}

In Fig.~\ref{fig:dsig_dW} we show differential distributions in the invariant
mass of the four-muon system (left panel) as well as distributions in the
transverse momentum of individual muons (right panel). 
The distributions in the left panel are done for different cuts 
on the transverse momenta of each of the muons.
There is a strong dependence on the cuts. However, the cross sections
for double-scattering are considerably larger than those for
single scattering. We conclude that four-muon
events measured in a future will originate mostly from the double-scattering mechanism,
which is easier to handle in the case of nuclear calculations. 
In the right panel we show the transverse momentum distributions of a muon 
for four-muon events (bottom curve) and for two-muon events (top
curve). The two
distributions have the same shape in transverse momentum.
The distribution in transverse momentum is very steep which explains
the strong dependence on the transverse momentum cut as demonstrated 
e.g.\ in the left panel.

\begin{figure}
\includegraphics[scale=0.4]{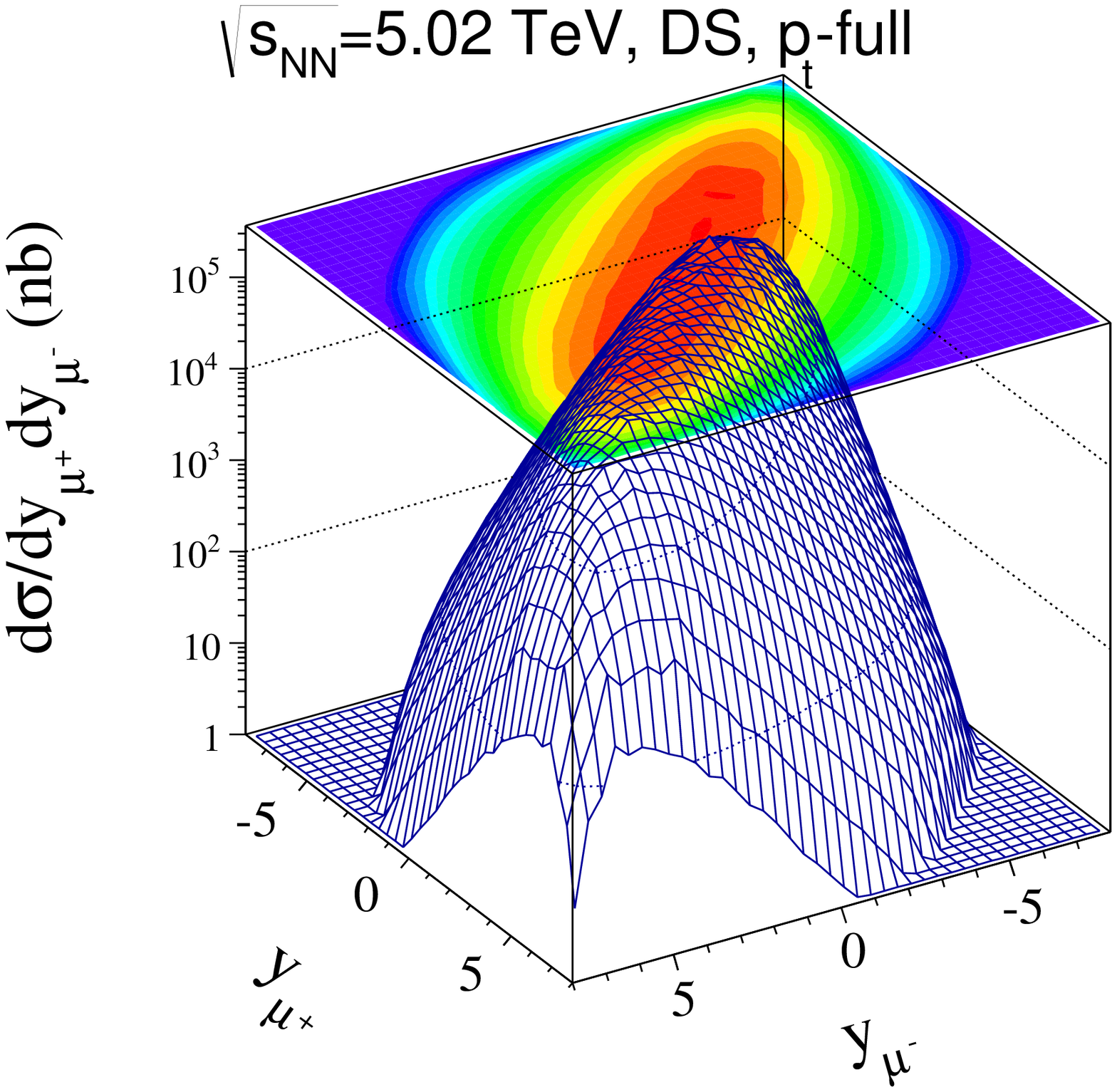}
\includegraphics[scale=0.4]{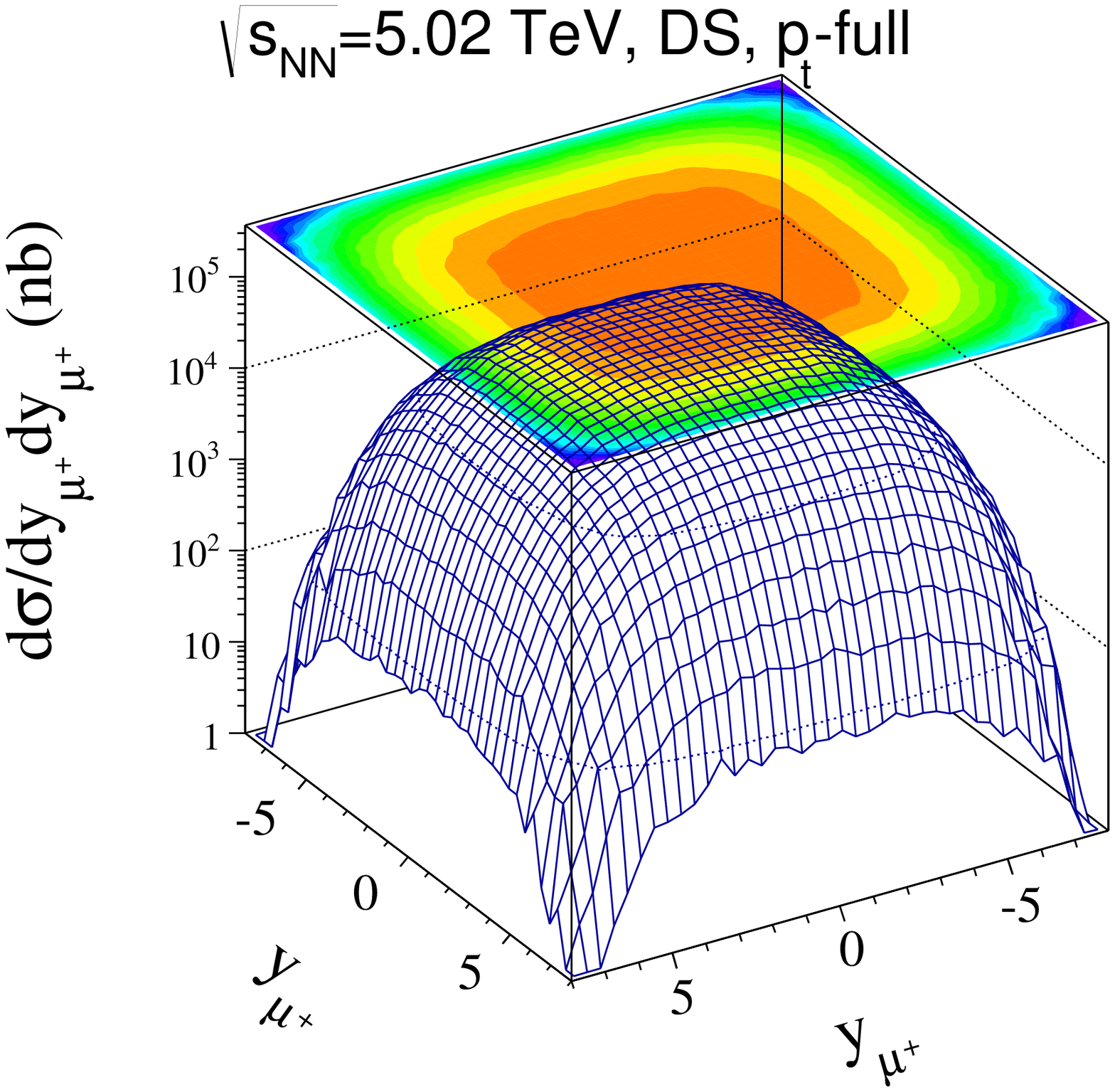}
\caption{Two-dimensional cross section as function of the rapidities of 
muons for the double-scattering mechanism.
The left panel is for $d\sigma/dy_{\mu^+}dy_{\mu^-}$ and the
right panel is for 
$d\sigma/dy_{\mu^+}dy_{\mu^+} = d\sigma/dy_{\mu^-}dy_{\mu^-}$.
No extra cuts on rapidities or transverse momenta were imposed here.}
\label{fig:sig_y1y2}
\end{figure}

Finally we wish to show some two-dimensional distributions.
We start with distributions in rapidities of two of the four muons
in the final state, see Fig.~\ref{fig:sig_y1y2}. The left panel is for
the opposite-sign muons from the same scattering, while the right panel
is for the same-sign muons, originating evidently from different
scatterings. There is also a similar distribution for opposite sign
muons originating from different scatterings. 
In a realistic experiment one cannot exactly determine from which 
scatterings the two randomly chosen muons are. However, muons coming from the same 
scattering have practically the same transverse momenta. This (transverse momentum balance) can be used 
to construct experimental distributions similar to those presented here.
Real experiments have, however, a limited range of muon rapidities.
In principle, ATLAS and CMS could try to construct such two-dimensional 
distributions. However, we do not expect too large statistics (see Table~I below), so one-dimensional distributions in $y_{diff}$ may be the 
only practical option.

\begin{figure}
\includegraphics[scale=0.4]{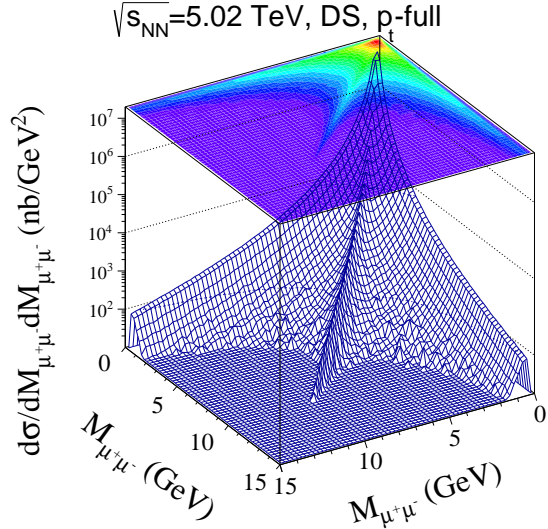}
\caption{Two-dimensional cross section as function of the invariant masses
of the first and second pair of muons in the double-scattering mechanism.
No extra cuts on rapidities or transverse momenta were imposed here.}
\label{fig:sig_MM}
\end{figure}

In Fig.~\ref{fig:sig_MM} we show two-dimensional distributions
in the invariant masses for the first and second scattering.
On the experimental side one would need to select opposite-sign muons in each
pair and use the transverse momentum balance check to identify the ``first'' and ``second'' 
scattering. We expect that there will be cases when such an identification
may not be possible due to finite transverse momentum resolution. 
The actual efficiency of such an identification
of the two scatterings requires dedicated Monte Carlo studies.
This goes beyond the scope of the present paper.

Finally, in Table~I we show predicted numbers of counts for
different transverse momentum and rapidity cuts specified in the table.
In this calculation we have assumed an integrated luminosity 
$L_{int} =$ 1 nb$^{-1}$.

\begin{table}[!h]
	\caption{The cross section in nb for selected cuts and number of counts
		for selected integrated luminosity $L_{int}$ = 1 nb$^{-1}$.}
\begin{tabular}{c|r|r}
\hline
experimental cuts      & cross section & number of counts \\               
\hline
-2.5 $< y_i <$ 2.5, $p_t >$ 0.5 GeV  & 815 	&  815 \\
-2.5 $< y_i <$ 2.5, $p_t >$ 1.0 GeV  &  53 	&   53 \\
-0.9 $< y_i <$ 0.9, $p_t >$ 0.5 GeV  &  31	&   31 \\
-0.9 $< y_i <$ 0.9, $p_t >$ 1.0 GeV  &   2 	&    2 \\
-2.4 $< y_i <$ 2.4, $p_t >$ 4.0 GeV  &   1.9$\times$E-3 	&    $\ll$1 \\
\hline
\end{tabular}
\end{table}

Rather small numbers of counts are predicted for the selected
luminosity $L =$ 1 nb$^{-1}$. For the cuts relevant for
the present ATLAS measurement of dimuons, $p_{t,cut}$ = 4 GeV,
it would be very difficult to observe any event with four muons.
For CMS and ALICE the situation seems to be better with $p_{t,cut}$ = 1 GeV.
A real measurement would be a first experimental confirmation
of double-scattering effects in UPC.

\section{Conclusions}

In the present paper we have discussed production of two pairs
of muons (four muons) in ultraperipheral ultrarelativistic heavy ion collisions
at $\sqrt{s_{NN}}$ = 5.02 TeV.
We have included both the double-scattering mechanism discussed
before for production of two electron pairs and the single-scattering mechanism for the first time.

We have presented several distributions for the elementary process
$\gamma \gamma \to \mu^+ \mu^- \mu^+ \mu^-$.
The calculation have been done using the automated code \KaTie\
adapted for the double-photon induced processes.
The cross section for different cuts on muon transverse momenta
has been calculated as a function of the (sub)collision energy.

The elementary cross sections have been calculated on a grid, and this
grid was used next for calculating nuclear $Pb Pb \to Pb Pb 4 \mu$ cross sections.
The flux of photons has been calculated using a realistic nuclear charge 
form factor, being a Fourier transform of the realistic charge distribution.
Several differential single-particle distributions and 
correlation distributions have been calculated and presented.
For a first time, we have shown explicitly that the cross section 
for the single-scattering mechanism is considerably smaller than the cross section
for the double-scattering mechanism. This shows that the
double-scattering mechanism is sufficient for detailed studies
and planning experiments.

We have shown a few one- (rapidity, transverse momentum, invariant
masses) and some two-dimensional distributions
that could be measured at the LHC.
The cross sections for four-muon production strongly depend on 
the cuts on muon transverse momenta.
We have discussed how to study some correlation observables
from the same $\gamma \gamma$ scattering by using the transverse momentum 
balance of opposite-sign muons.

Finally, we have presented counting rates for experimental situations,
i.e.\ including cuts on transverse momenta and rapidities.
Measurable cross sections and counting rates of the order of hundreds of events have been predicted.
We hope that some LHC collaborations (CMS and ALICE) will 
be able to start such studies in the near future
while for the present ATLAS cuts the situation seems rather difficult.

\vspace{1.5cm}

{\bf Acknowledments}

This work was partially supported by the Polish grant 
No. DEC-2014/15/B/ST2/02528 (OPUS)
as well as by the Centre for Innovation and Transfer of Natural Sciences
and Engineering Knowledge in Rzesz\'ow.
AvH was supported by grant of National Science Center, Poland, No. 2015/17/B/ST2/01838.
We are indebted to Peter Steinberg for a discussion on dimuon production related to recent ATLAS measurement.

\bibliography{refs}

\end{document}